\documentclass[modern]{aastex631}

\newcommand{\red}{\textcolor{red}}

\usepackage{amsmath}

\shorttitle{Electron acceleration and transport}
\shortauthors{Kong et al.}

\begin{document}

\title{Numerical Modeling of Energetic Electron Acceleration, Transport, and Emission in Solar Flares: Connecting Loop-top and Footpoint Hard X-Ray Sources}

\correspondingauthor{Xiangliang Kong}
\email{kongx@sdu.edu.cn}

\author[0000-0003-1034-5857]{Xiangliang Kong}
\affiliation{School of Space Science and Physics, Institute of Space Sciences, Shandong University, Weihai, Shandong 264209, People's Republic of China}
\affiliation{Institute of Frontier and Interdisciplinary Science, Shandong University, Qingdao, Shandong 266237, People's Republic of China}

\author[0000-0002-0660-3350]{Bin Chen}
\affiliation{Center for Solar-Terrestrial Research, New Jersey Institute of Technology, Newark, NJ 07102, USA}

\author[0000-0003-4315-3755]{Fan Guo}
\affiliation{Los Alamos National Laboratory, Los Alamos, NM 87545, USA}
\affiliation{New Mexico Consortium, Los Alamos, NM 87544, USA}

\author[0000-0002-9258-4490]{Chengcai Shen}
\affiliation{Center for Astrophysics, Harvard $\&$ Smithsonian, Cambridge, MA 02138, USA}

\author[0000-0001-5278-8029]{Xiaocan Li}
\affiliation{Department of Physics and Astronomy, Dartmouth College, Hanover, NH 03755, USA}

\author[0000-0002-5983-104X]{Jing Ye}
\affiliation{Yunnan Observatories, Chinese Academy of Sciences, Kunming, Yunnan 650216, People's Republic of China}

\author[0000-0003-3936-5288]{Lulu Zhao}
\affiliation{Department of Climate and Space Sciences and Engineering, University of Michigan, Ann Arbor, MI 48109, USA}

\author{Zelong Jiang}
\affiliation{School of Space Science and Physics, Institute of Space Sciences, Shandong University, Weihai, Shandong 264209, People's Republic of China}
\author[0000-0002-0660-3350]{Sijie Yu}
\affiliation{Center for Solar-Terrestrial Research, New Jersey Institute of Technology, Newark, NJ 07102, USA}

\author[0000-0001-6449-8838]{Yao Chen}
\affiliation{School of Space Science and Physics, Institute of Space Sciences, Shandong University, Weihai, Shandong 264209, People's Republic of China}
\affiliation{Institute of Frontier and Interdisciplinary Science, Shandong University, Qingdao, Shandong 266237, People's Republic of China}

\author[0000-0002-0850-4233]{Joe Giacalone}
\affiliation{Department of Planetary Sciences, University of Arizona, Tucson, AZ 85721, USA}



\begin{abstract}
The acceleration and transport of energetic electrons during solar flares is one of the outstanding topics in solar physics.
Recent X-ray and radio imaging and spectroscopy observations have provided diagnostics of the distribution of nonthermal electrons and suggested that, in certain flare events, electrons are primarily accelerated in the loop-top and likely experience trapping and/or scattering effects.
By combining the focused particle transport equation with magnetohydrodynamic (MHD) simulations of solar flares, we present a macroscopic particle model that naturally incorporates electron acceleration and transport.
Our simulation results indicate that the physical processes such as turbulent pitch-angle scattering can have important impacts on both electron acceleration in the loop-top and transport in the flare loop, and their influences are highly energy dependent.
A spatial-dependent turbulent scattering with enhancement in the loop-top can enable both efficient electron acceleration to high energies and transport of abundant electrons to the footpoints.
We further generate spatially resolved synthetic hard X-ray (HXR) emission images and spectra, revealing both the loop-top and footpoint HXR sources. Similar to the observations, we show that the footpoint HXR sources are brighter and harder than the loop-top HXR source.
We suggest that the macroscopic particle model provides new insights into understanding the connection between the observed loop-top and footpoint nonthermal emission sources by combining the particle model with dynamically evolving MHD simulations of solar flares.

\end{abstract}



\section{Introduction} \label{sec:intro}

Particle acceleration, transport, and subsequent emission processes are at the heart of the high-energy aspects of solar flares. Observations have suggested an enormous number of particles are accelerated to high energies and the nonthermal particles can carry a substantial portion ($\sim$10\%$-$50\%) of the released magnetic energy \citep{1976SoPh...50..153L,2012ApJ...759...71E,2017ApJ...836...17A,2020A&A...644A.172W}. These accelerated particles further propagate and precipitate, producing hard X-ray (HXR) footpoint sources in the dense chromosphere via thick-target bremsstrahlung and leading to chromosphere evaporation. Despite a long history of study, the whole process of energetic particles and their effects on flare dynamics is still an active field of research.

Flare-accelerated electrons produce nonthermal emissions in HXR and microwave wavelengths via the bremsstrahlung and gyrosynchrotron radiation mechanisms, respectively. Therefore, HXR and microwave emissions serve as important diagnostics for flare-accelerated nonthermal electrons \citep[see reviews by][]{2011SSRv..159..107H,2011SSRv..159..301K,2011SSRv..159..225W}.
 Nonthermal emission sources have been frequently observed at or above the top of extreme-ultraviolet (EUV) or soft X-ray (SXR) flare loops \citep[e.g.,][]{1994Natur.371..495M,2002ApJ...580L.185M,2002ApJ...569..459P,2012ApJ...748...33C,2013ApJ...767..168L,2013A&A...551A.135S,2013NatPh...9..489S,2014ApJ...780..107K,2015SoPh..290...79K,2015ApJ...799..129O,2017ApJ...843....8C,2020ApJ...900...17Y,2022ApJ...924L...7L}.
The location of the observed loop-top nonthermal sources indicates that the primary particle acceleration may take place in the corona, probably close to the emission source itself (e.g., by turbulence or shocks). \citet{2020NatAs...4.1140C} measured the spatial distribution of magnetic field and microwave-emitting relativistic electrons along a large-scale current sheet in the 2017 September 10 flare and found that the loop-top region with a local minimum of magnetic field (referred to as a ``magnetic bottle") coincided with the location where most of the high-energy electrons were present. They suggested that the loop-top magnetic bottle may be the primary site to accelerate and/or confine energetic electrons. In the same flare but for the main impulsive phase, \citet{2022Natur.606..674F} revealed a volume in the loop-top filled with almost only nonthermal ($>$20 keV) electrons and suggested that a large fraction of electrons there experienced a prominent acceleration.

Recent modeling effort has been successful in modeling particle acceleration in the loop-top and current sheet regions \citep[e.g.,][]{2019ApJ...887L..37K,2021PhRvL.126m5101A,2022ApJ...932...92L}. Although multiple acceleration mechanisms may be relevant for flare particle acceleration \citep[see, e.g.,][]{1997JGR...10214631M,2011SSRv..159..357Z, 2021PhPl...28e2905L}, in this study, we focus our discussion on the flare termination shock (TS), which is capable of directly accelerating particles in the loop-top region and producing loop-top emissions \citep{1998ApJ...495L..67T,2009A&A...494..669M,2009A&A...494..677W,2012ApJ...753...28G,2013ApJ...769...22L,2015Sci...350.1238C,2019ApJ...887L..37K}.
In the standard model of solar flares, this TS forms when the reconnection outflows impinge upon the top of flare arcades, serving as one promising acceleration mechanism in the loop-top region \citep{1994Natur.371..495M,1995ApJ...451L..83S}.
The flare TS has long been predicted in 2D/2.5D magnetohydrodynamic (MHD) simulations \citep[e.g.,][]{1986ApJ...305..553F,1996ApJ...466.1054M,2015ApJ...805..135T,2017ApJ...848..102T,2018ApJ...869..116S,2019MNRAS.489.3183C,2019MNRAS.482..588Y,2020ApJ...896...97R,2020ApJ...898...90Z,2021ApJ...923..227W} and lately in the 3D MHD model as well \citep[e.g.,][]{2022NatAs...6..317S}. \footnote{Regarding the observational evidence of the TS, a handful of events have been reported \citep[e.g.,][]{2002A&A...384..273A,2004ApJ...615..526A,2009A&A...494..669M,2009A&A...494..677W,2015Sci...350.1238C,2018ApJ...865..161P,2021ApJ...911....4L,2022ApJ...929...99C}.
We refer the readers to \citet{2019ApJ...884...63C} for more discussions on various observational signatures of TSs and their detectability.}
Recently, \citet{2019ApJ...887L..37K} presented macroscopic numerical modeling of electron acceleration by the TS by coupling the Parker transport equation \citep{1965P&SS...13....9P} with an MHD simulation of a classic two-ribbon flare. They showed that electrons are mainly accelerated at the TS and concentrated in the loop-top, and a magnetic trap in the loop-top plays an important role in both accelerating and confining electrons.
\citet{2022ApJ...933...93K} further suggested that the TS acceleration mechanism can also explain the double coronal HXR sources as observed in some solar flares \citep[e.g.,][]{2012ApJ...748...33C}.

To understand the connection between emissions at the loop-top and footpoints, one needs to study how electrons propagate and precipitate to the footpoints, and further produce nonthermal emissions. Spatially resolved X-ray imaging spectroscopy from \textit{RHESSI} has provided the opportunity to study the coronal and footpoint HXR sources in a solar flare simultaneously. \citet{2013A&A...551A.135S} revealed that the nonthermal electron rate (in electrons s$^{-1}$) in the loop-top source is significantly larger than that in the footpoint sources (by a factor of $\sim$2$-$8). This observational result suggests that the energetic electrons experience significant trapping in the coronal part of the flare loop (or in the above-the-looptop region), and they are not free-streaming and should be subject to transport effects.
By assuming a single power-law electron spectrum and applying thin-target and thick-target bremsstrahlung models for the loop-top and footpoint sources, they deduced the corresponding electron spectral indices and found the loop-top spectral indexes are smaller than the footpoint indexes by 0.2--1, possibly implying a softening in the electron spectrum.
Earlier studies have also shown that the HXR spectral indices between loop-top and footpoint sources can differ significantly from 2 \citep[e.g.,][]{2002ApJ...569..459P,2006A&A...456..751B}, which is the value expected from the thin- and thick-target bremsstrahlung provided that the emissions arise from the same population of nonthermal electrons with a single power-law spectrum \citep{2018SSRv..214...82O}.
In addition to the softening in electron spectrum at the footpoints, other scenarios may also explain the difference in photon spectral index being smaller than 2, including, e.g., the electron spectrum deviating from a single power-law with a break or rollover at high energies \citep{2003ApJ...586..606H,2011SSRv..159..107H}, the loop-top source not being completely thin-target (e.g., thick-target for low energy photons; \citealt{2006A&A...456..751B}). 
On the other hand, the difference in photon spectral index being larger than 2 may reflect the hardening in electron spectrum due to transport effects such as Coulomb collisions and return current, which can cause low-energy electrons to preferentially lose their energies \citep{2006A&A...456..751B,2017ApJ...851...78A}. In addition, albedo effects can modify the HXR photon spectrum at the footpoints, causing deviations from the ideal thin--thick-target relation \citep[e.g.,][]{2006A&A...446.1157K}.

In previous flare models, particle acceleration and transport effects are usually treated separately. Various transport processes have been considered, including magnetic mirroring due to magnetic field convergence, energy loss and pitch-angle scattering due to Coulomb collisions with the ambient plasma, pitch-angle scattering by magnetic turbulence, and return current \citep[e.g.,][]{1998ApJ...505..418F,2000ApJ...543..457L,2006ApJ...647.1472K,2006ApJ...651..553Z,2011ApJ...732..111M,2012ApJ...752....4B,2013ApJ...777...33C,2014ApJ...787...86J,2014ApJ...780..176K,2017ApJ...835..262B,2018ApJ...868L..28E,2018A&A...610A...6M,2020ApJ...902...16A,2020ApJ...904....1T}.
These transport effects mostly operate on different time scales and the interplay between them can significantly modify the spatial and energy spectral distributions of flare-accelerated electrons.
For example, by comparing with the observations in a flare event, \citet{2018A&A...610A...6M} showed that the turbulent pitch-angle scattering can explain the coronal trapping of energetic electrons and the spectral hardening between the loop-top and footpoints.
Although the transport of nonthermal electrons in the flare loop has been intensively studied, most previous models used a simplified 1D or semicircular coronal loop and the properties of injected electrons (e.g., energy spectrum, pitch-angle distribution, spatial extent and time profile) are based on assumptions.

Until now, there has been a lack of realistic models that incorporate both particle acceleration and transport in the flare region.
To connect the emission sources in the loop-top and at the footpoints, it is critical to develop a model for investigating both the acceleration and transport processes, and further predict radiation signatures. In the loop-top regions, the primary particle acceleration mechanism, either stochastic or shock acceleration, requires sufficient trapping of particles in the acceleration site \citep[see reviews,][]{2012SSRv..173..535P,2021FrASS...8...27G}.
This means that the transport effects not only affect the escape of electrons from the loop-top, but also affect the rate of particle acceleration.
In this paper, we present a macroscopic particle model by coupling the focused particle transport equation with an MHD simulation of the solar flare. This particle model naturally incorporates both the acceleration of nonthermal electrons in the loop-top and the transport of electrons in the flare loop. We focus on the scenario that a flare TS forms and accelerates electrons at the loop-top, and the accelerated electrons further precipitate to the footpoints. We find that physical processes such as turbulent scattering can have important impacts on both the electron acceleration in the loop-top and the subsequent transport in the flare loop, and the influences are highly energy-dependent. We further calculate spatially resolved synthetic HXR emission images and spectra, revealing that the footpoint sources are brighter and harder than the loop-top source, as expected.
In Section 2, we describe the numerical methods. In Section 3, we present the simulation results, with an emphasis on the effects of turbulent pitch-angle diffusion on the spatial distribution and energy spectra of nonthermal electrons. Spatially resolved synthetic HXR images and spectra that can be directly compared with observations are generated and discussed. Conclusions and discussion are presented in Section 4.

\section{Numerical Methods} \label{sec:methods}

\subsection{MHD simulation of the solar flare}

We first perform an MHD simulation of a classic two-ribbon solar flare by numerically solving the 2.5D resistive MHD equations using the Athena MHD code \citep{2008ApJS..178..137S}. Detailed discussion of the model setup can be found in  \citet{2018ApJ...869..116S} and here we only provide a salient
description. The initial setup is a force-free current sheet along the $y$ direction (height) with a uniform guide field $B_g$ = 0.1 $B_0$, where $B_0$ is the normalized magnetic field.
To achieve the two-ribbon flare configuration, the magnetic field lines at the bottom boundary are set to be line tied on the photosphere.
We include classical Spitzer thermal conduction and the background plasma beta is $\beta_0$ = 0.01.
We use a uniform resistivity corresponding to a constant magnetic Reynolds number $R_m$ = $5 \times 10^{4}$.
The simulation domain is $x$ = [$-$1, 1] and $y$ = [0, 2]. We use uniform grid and the grid numbers are $N_x \times N_y$ = 1155 $\times$ 1155. The simulation results are normalized by the length $L_0$ = 75 Mm, the plasma density $\rho_0 = 1.93 \times 10^{-12}$ kg m$^{-3}$ (number density $n_0 = 1.153 \times 10^{9}$ cm$^{-3}$), and the magnetic field strength $B_0$ = 40 G. This gives the characteristic Alfv$\acute{e}$n speed is $V_0$ = 2569 km s$^{-1}$, and a characteristic time $t_0$ = $L_0/V_0$ = 29.2 s.

\begin{figure}
\centering
\includegraphics[width=0.95\linewidth]{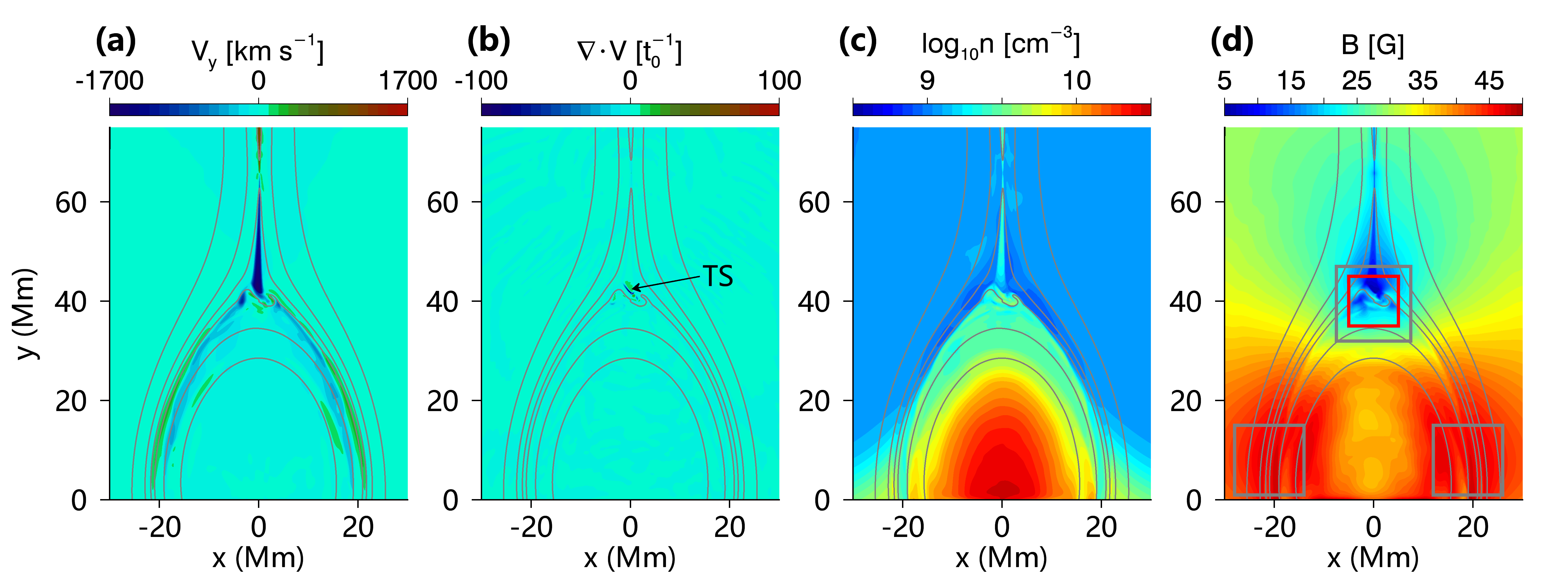}
\caption{
Spatial distributions of MHD parameters at the MHD simulation time 92 $t_0$: (a) the plasma velocity in $y$ direction ($V_y$), (b) the divergence of plasma velocity ($\nabla \cdot \textbf{V}$) with the TS marked by the arrow, (c) the plasma number density ($n$), and (d) the magnetic field strength ($B$).
Gray curves in each panel indicate the magnetic field lines. In panel (d), the red box indicates the loop-top region where the turbulent scattering is enhanced in Run S and Run S-loss.
}
\label{fig:mhd}
\end{figure}

Figure \ref{fig:mhd} shows the spatial distributions of MHD parameters at the simulation time 92 $t_0$, including the plasma flow velocity in the $y$ direction ($V_y$), the divergence of flow velocity ($\nabla \cdot \textbf{V}$), the plasma number density ($n$), and the magnitude of magnetic field ($B$).
A TS forms in the loop-top where the downward reconnection flow encounters closed magnetic loops. It is manifested by negative $\nabla \cdot \textbf{V}$ owing to strong compression.
As shown in Figure \ref{fig:mhd}(d), the magnetic field strength is weaker in the loop-top and current sheet regions.
We measure the strength of magnetic field averaged over the three gray boxes in Figure \ref{fig:mhd}(d) and find that it is $\sim$21 G in the loop-top and $\sim$46 G in the two footpoints.
Therefore, the magnetic mirror ratio in the flare loop along which most electrons gyrate is $\sim$2.2.
As shown in previous flare transport models, magnetic mirroring can play an important role in the trapping of electrons in the loop-top \citep[e.g.,][]{1998ApJ...505..418F,2012ApJ...752....4B}.
We will discuss the effect of magnetic mirroring on particle acceleration and transport in our future work.

\subsection{Particle acceleration and transport model}

The most fundamental description of the motion of charged particles is the Newton-Lorentz equation. However, because the gyroradius of electrons in the low corona ($\sim$0.01$-$1 m) is much smaller than the macroscopic flare scale ($\sim$10$^8$ m), it is computationally not feasible to follow the full electron trajectories. 
One practical approach is to trace the gyro-centers of electrons instead, so-called the guiding center approximation \citep{1963RvGSP...1..283N}.
The guiding center approach combined with MHD simulations has been used to study particle acceleration and transport in current sheet reconnection and flares
\citep[e.g.,][]{2006ApJ...647.1472K,2010ApJ...720.1603G,2020ApJ...902..147G,2015RAA....15..348Y,2015ApJ...815....6Z}. More recently, a new model $kglobal$ was developed, which includes the feedback from energetic electrons to the MHD flow dynamics \citep[e.g.,][]{2021PhRvL.126m5101A}. While in principle one can include effects of the interaction between turbulence/waves and particles, the standard version of the guiding-center approach assumes an adiabatic process.

In general, the standard approach for studying energetic particle acceleration and transport is to use particle transport theory \citep{2014LNP...877.....Z}.
For charged particles experiencing strong scattering in a turbulent magnetized plasma, the evolution of particle distribution function can be described by the Parker transport equation \citep{1965P&SS...13....9P}.
Parker equation is a convective-diffusive equation including the effects of convection, diffusion, drift, and acceleration, and assumes a nearly isotropic pitch-angle distribution.
By coupling with MHD simulations, it has been applied to modeling electron acceleration by the flare TS in the loop-top \citep{2019ApJ...887L..37K,2020ApJ...905L..16K,2022ApJ...933...93K} and by large-scale compression in the reonnection layer \citep{2018ApJ...866....4L,2022ApJ...932...92L}.
However, in the context when the anisotropy is large, one should use the focused transport equation, which retains the pitch-angle dependence of the distribution function \citep[see the review,][]{2020SSRv..216..146V}.
In addition to similar terms in the Parker equation, the focused transport equation contains other terms, e.g., streaming along the magnetic field and variation of pitch-angle.
It has been widely applied to the transport of solar energetic particles (SEPs) in the corona and interplanetary space
\citep[e.g.,][]{2012ApJ...752...37W,2016ApJ...821...62Z,2017JGRA..12210938H,2017ApJ...846..107Z,2019JASTP.182..155W,2019A&A...622A..28W} and particle acceleration at shocks \citep[e.g.,][]{2012ApJ...746..104L,2011ApJ...738..168Z,2013ApJ...767....6Z,2016ApJ...820...24K}.

In this work, we use the focused transport equation to model particle acceleration and transport in solar flares.
The basic equation  can be written as \citep[e.g.,][]{2006JGRA..111.8101Q,2009ApJ...692..109Z,2011ApJ...738..168Z,2017ApJ...846..107Z}:
\begin{eqnarray}
\frac{\partial f}{\partial t} & = & \nabla \cdot \boldsymbol{\kappa}_{\perp} \cdot \nabla f - (v \mu \hat{\boldsymbol{b}} + \boldsymbol{U} + \boldsymbol{V_d}) \cdot \nabla f  \nonumber \\
  & + & \frac{\partial}{\partial \mu} D_{\mu\mu} \frac{\partial f}{\partial \mu} - \frac{d \mu}{dt} \frac{\partial f}{\partial \mu} - \frac{dp}{dt} \frac{\partial f}{\partial p},
\end{eqnarray}
where $f(\boldsymbol{X},\mu,p,t)$ is the gyrophase-averaged distribution function of charged particles as a function of spatial location $\boldsymbol{X}$, momentum $p$, pitch-angle cosine $\mu$, and time $t$.
The terms on the right-hand side contain cross-field spatial diffusion with a tensor $\boldsymbol{\kappa}_{\perp}$, streaming along the ambient magnetic field direction $\hat{\boldsymbol{b}}$ with particle speed $v$, advection with the background plasma $\boldsymbol{U}$, magnetic gradient or curvature drift $\boldsymbol{V_d}$, pitch-angle diffusion with a coefficient $D_{\mu\mu}$, focusing $d\mu/dt$, and adiabatic cooling/gain $dp/dt$. Note that the momentum diffusion term is not included in Equation (1).

In the adiabatic approximation, the pitch-angle change and momentum change terms can be calculated from the magnetic field $\boldsymbol{B} = B \hat{\boldsymbol{b}}$ and plasma velocity $\boldsymbol{U}$:
\begin{eqnarray}
\frac{d \mu}{dt} & = & \frac{1-\mu^2}{2} \left[-\frac{v}{L_B} + \mu(\nabla \cdot \boldsymbol{U} - 3\hat{\boldsymbol{b}}\hat{\boldsymbol{b}}: \nabla \boldsymbol{U}) \right], \\
\frac{dp}{dt} & = &  -p \left[ \frac{1-\mu^2}{2} (\nabla \cdot \boldsymbol{U} - \hat{\boldsymbol{b}}\hat{\boldsymbol{b}}: \nabla \boldsymbol{U}) + \mu^2 \hat{\boldsymbol{b}}\hat{\boldsymbol{b}}: \nabla \boldsymbol{U}  \right].
\end{eqnarray}
The pitch angle change contains magnetic mirroring effect with a scale length $L_B = (\hat{\boldsymbol{b}} \cdot \nabla ln B)^{-1}$ describing the gradient in the magnetic field direction. The momentum change is related to significant compression acceleration at the shock where the divergence in the plasma flow velocity is negative, as in the Parker transport equation, and also incompressible shear effects \citep[e.g.,][]{2012ApJ...746..104L,2018ApJ...855...80L}.

We use the stochastic integration approach to numerically solve the focused transport equation. Because the transport equation is essentially a Fokker-Planck equation, it is mathematically equivalent to a set of time-forward stochastic differential equations (SDEs) \citep[e.g.,][]{1999ApJ...513..409Z,2012CoPhC.183..530K,2017SSRv..212..151S,2017ApJ...846..107Z}:
\begin{eqnarray}
d\boldsymbol{X} & = &  (v \mu \hat{\boldsymbol{b}} + \boldsymbol{U} + \nabla \cdot \boldsymbol{\kappa}_{\perp})dt + \sqrt{2\boldsymbol{\kappa}_{\perp}} \cdot d\boldsymbol{W}_x(t), \\
dp & = & \frac{dp}{dt}dt, \\
d\mu & = & \left[\frac{d\mu}{dt} + \frac{\partial D_{\mu\mu}}{\partial \mu} \right]dt + \sqrt{2D_{\mu\mu}}dW_\mu(t),
\end{eqnarray}
where $d\boldsymbol{W}_x$ and $dW_\mu$ are Wiener processes. Note that the drift term $\boldsymbol{V_d}$ is not considered in Equation (4) for our 2D simulations.

In addition to the plasma velocity and magnetic field from MHD simulations, we need to specify the diffusion coefficients, including the perpendicular spatial diffusion coefficient $\kappa_{\perp}$ and the pitch-angle diffusion coefficient $D_{\mu \mu}$.

In the quasi-linear theory, the resonant interaction between the particle and the turbulent magnetic field can be related by the pitch-angle diffusion coefficient $D_{\mu \mu}$ \citep{1971RvGSP...9...27J},
\begin{equation}
D_{\mu \mu} = \frac{\pi}{4} \Omega_0 (1-\mu^2) \frac{k_r P(k_r)}{B_0^2},
\end{equation}
where $\Omega_0 = qB_0/m$ is the particle gyrofrequency with the mass $m$ and the charge $q$, $P(k)$ is the turbulence power spectrum, and $k_r=\Omega_0/(v|\mu|)$ is the resonant wavenumber.

We assume the turbulence power spectrum $P(k)$ in the form of
\begin{equation}
P(k) = A_0 L_c \sigma^2 B_0^2 \frac{1}{1 + (k L_c)^{\Gamma}},
\end{equation}
\noindent where $k$ is the wave number, $L_c$ is the turbulence correlation length, $\sigma^2$ = $\langle \delta B^2\rangle / B_0^2$ is the variance of turbulent magnetic field, $\Gamma$ is the spectral index, and $A_0$ is the normalization constant. For the Kolmogorov spectrum with $\Gamma$ = 5/3, $A_0 = 5/(3\pi)sin(3\pi/5) \approx$ 0.5.

In the non-relativistic limit,  we take the pitch-angle diffusion coefficient in the form of
\begin{equation}
    D_{\mu \mu} = D_{\mu \mu 0} \left(\frac{p}{p_0}\right)^{\Gamma-1} (1-\mu^2)(|\mu|^{\Gamma -1} + h_0), 
\end{equation}
where $D_{\mu \mu 0} = \frac{\pi}{4}  A_0 \sigma^2 \Omega_0^{2-\Gamma} L_c^{1-\Gamma} v_0^{\Gamma-1}$, $p_0$ ($v_0$) is the initial particle momentum (velocity) at the injection energy ($E_0$ = 5 keV). The parameter $h_0$ is added to describe the scattering through $\mu$ = 0 and we set $h_0$ = 0.2 \citep[e.g.,][]{2017ApJ...846..107Z}.
For the reference run (Run S as listed in Table 1), we assume $L_c$ = 1 Mm, $B_0$ = 40 G, $\sigma^2$ = 0.05, therefore $D_{\mu \mu 0}$ = 288 s$^{-1}$ for 5 keV electrons.

The spatial diffusion coefficient along the direction of the magnetic field can be related to the pitch-angle diffusion coefficient by \citep[e.g.,][]{1966ApJ...146..480J,1976JGR....81.2089L}
\begin{equation}
\kappa_\parallel (v) = \frac{v^2}{4} \int_0^1 \frac{(1 - \mu^2)^2}{D_{\mu \mu}} d\mu.
\end{equation}

By substituting Equations (7) and (8) into Equation (10), we obtain the parallel diffusion coefficient \citep[e.g.,][]{1999ApJ...520..204G,2022ApJ...925L..13Y}:
\begin{equation}
    \kappa_\parallel= \frac{v^3}{A_0 \pi L_c \Omega_0^2 \sigma^2}
    \left[\frac{1}{4}+\left(\frac{\Omega_0L_c}{v}\right)^{\Gamma}\frac{2}{\left(2-\Gamma\right)
    \left(4-\Gamma\right)}\right].
\end{equation}

For the Kolmogorov spectrum with $\Gamma$ = 5/3, $\kappa_\parallel \approx 1.62 v^{4/3} L_c^{2/3} \Omega_0^{-1/3} / \sigma^2$ \citep{2022ApJ...932...92L}.
For the reference run (Run 1), $\kappa_{\parallel 0}$ = 3.91 $\times$ 10$^{12}$ m$^2$ s$ ^{-1}$ = 0.02 $\kappa_0$ for 5 keV electrons, where the normalization $\kappa_0 = L_0 V_0$ = 1.93 $\times$ 10$^{14}$ m$^2$ s$^{-1}$.

For the perpendicular diffusion coefficient $\kappa_\perp$, we take $\kappa_\perp/\kappa_\parallel$ = 0.01 in Run 1, similar to results of test-particle simulations in synthetic turbulence \citep{1999ApJ...520..204G}.
We consider $\kappa_\perp$ in the form of $\kappa_\perp = \kappa_{\perp 0} (p/p_0)^{4/3}$,
where $\kappa_{\perp 0}$ = 0.01 $\kappa_{\parallel 0}$ = 2.03 $\times$ 10$^{-4}$ $\kappa_0$.
Perpendicular diffusion can affect electron acceleration in the loop-top and the size of X-ray sources \citep[][]{2011A&A...535A..18B,2011ApJ...730L..22K}. For simplicity, we assume the same $\kappa_{\perp 0}$ in all simulations. Note that the ratio of $\kappa_\perp/\kappa_\parallel$ is reduced in the case of weaker turbulent scattering.

For the transport of SEPs in the interplanetary space, particle-particle collisions are not important because the interplanetary space is extremely tenuous. However, in the context of solar flares in the low corona, Coulomb collisions between energetic electrons and the ambient plasma should be considered due to high plasma density. The effects of collisions are two-fold, i.e., energy loss and pitch-angle scattering.

The collisional energy loss rate in non-relativistic limit is \citep{1971SoPh...18..489B,1978ApJ...224..241E,2011SSRv..159..107H}
\begin{equation}
\frac{dE}{dt} = - \frac{K}{E} n_{th} v,
\end{equation}
where $E$ is the electron energy, $v$ is the electron speed (cm s$^{-1}$), $K = 2 \pi e^4 \Lambda$, $e$ is the electronic charge (e.s.u.), $\Lambda$ is the Coulomb logarithm, $n_{th}$ is the number density (cm$^{-3}$) of thermal electrons. Because $\Lambda$ typically falls in the range of 20-30 for X-ray emitting electrons, the collisional parameter $K$ can be taken as a constant. With $dE/dt$ in keV s$^{-1}$ and $E$ in keV, the expression of $K$ can be written as \citep{2011SSRv..159..107H}
\begin{equation}
K = 3.0 \times 10^{-18} \left(\frac{\Lambda}{23}\right)  ~\rm keV^2 cm^2.
\end{equation}
As shown in Equation (12), the collisional energy loss rate is most significant for low-energy electrons. Therefore, it can cause the hardening of energy spectrum in the low-energy portion as energetic electrons stream downward from the loop-top to the footpoints.

In non-relativistic limit, $E = p^2 / 2 m_e$, where $m_e$ is the electron mass,  the extra term that should be added to the momentum change (Equation (3)) is given by,
\begin{equation}
\frac{dp}{dt} = - \frac{2 K m_e n_{th}}{p^2}.
\end{equation}

Coulomb collisions can also contribute to pitch-angle scattering. Considering fully ionized plasma and taking account of electron–electron and electron–hydrogen scattering, the collisional pitch-angle diffusion coefficient is \citep[e.g.,][]{1998ApJ...505..418F,2014ApJ...780..176K}
\begin{equation}
D_{\mu \mu}^{C} = \frac{2K n_{th}}{m_e^2 v^3} (1 - \mu^2).
\end{equation}
For 5 keV electron, by assuming the thermal plamsa density $n_{th}$ = $n_0$ = 1.153 $\times 10^{9}$ cm$^{-3}$ and $\Lambda$ = 23, we can calculate $D_{\mu \mu 0} ^{C}$ = $2K n_{th} / m_e^2 v^3$ = 0.29 s$^{-1}$. Considering the turbulent diffusion coefficient $D_{\mu \mu 0}$ = 288 s$^{-1}$ as assumed above, the collisional pitch-angle scattering rate is much less than the turbulent scattering rate, therefore the collisional pitch-angle scattering is negligible.

\subsection{Model coupling and simulation parameters}
The particle transport equation is coupled with the flare MHD simulation in a post-processing manner.
We numerically solve the SDEs (Equations 4$-$6) of the particle transport equation based on the time-dependent fluid velocity and magnetic field from MHD simulations.
The selected region for particle simulation is, $x$ = [$-$30, 30] Mm and $y$ = [0, 75] Mm (see Figure \ref{fig:mhd}).
We focus on a period between 91$-$92 $t_0$ in the MHD simulation when the loop-top region is relatively stable.
The temporal cadence of MHD frames is 0.005 $t_0$ (201 frames in total) and no interpolation is applied in time, meaning that we assume steady flow and magnetic fields between adjacent MHD frames. We use a bilinear interpolation in space to deduce the physical quantities and their partial derivatives at the particle position.

As noted above, the turbulence variance $\sigma^2$ = $\langle \delta B^2\rangle / B_0^2$ = 0.05 in the reference run, Run S. We calculate the mean free path $\lambda_{\parallel 0}$ = $3\kappa_{\parallel 0} / v_0$ = 2.8 $\times$ 10$^{5}$ m for 5 keV electrons.
From the observations of HXR sources in some flares, the mean free path was found to be in the order of 10$^{6}$-10$^{7}$ m for $\sim$30 keV electrons \citep[e.g.,][]{2014ApJ...780..176K,2018A&A...610A...6M}.
To examine the effects of turbulent pitch-angle scattering on particle acceleration and transport, we also consider the case with a weaker turbulent scattering. As listed in Table 1, in Run W, $D_{\mu \mu 0}$ is reduced to 57.7 s$^{-1}$, corresponding to $\sigma^2$ = 0.01 and $\lambda_{\parallel 0}$ = 1.4 $\times$ 10$^{6}$ m.

Turbulent pitch-angle scattering not only affects the transport of nonthermal electrons in the flare loop, but also affects electron acceleration by the TS in the loop-top. Electrons can be more efficiently accelerated when the shock propagates through large-scale turbulent magnetic field \citep{2021FrASS...8...27G}.
In observations, using nonthermal broading of spectral lines by \textit{Hinode}/EIS, it was suggested that the plasma turbulence is the highest in the loop-top \citep[e.g.,][]{2017PhRvL.118o5101K,2021ApJ...923...40S}.
MHD simulations also revealed that the loop-top is turbulent due to the impact of reconnection outflows and a variety of instabilities can develop \citep[e.g.,][]{2016ApJ...823..150T,2018ApJ...869..116S,2022NatAs...6..317S,2022ApJ...931L..32W}.
We consider spatial-dependent turbulent scattering in a simulation, i.e., Run SW as listed in Table 1. Compared with the weak scattering run (Run W), the turbulent scattering is enhanced in the loop-top region (as indicated by the red box in Figure \ref{fig:mhd}) with $D_{\mu \mu 0}$ (loop-top) = 288 s$^{-1}$, being the same value as in Run S.
For the corresponding simulations with collisional energy loss, they are named as Run S-loss, Run W-loss, and Run SW-loss, respectively, as shown in Table 1.

In all simulations, we assume an injection of 5 keV electrons with an isotropic pitch-angle distribution. In the 2017 September 10 flare, it was suggested that the plasma in the current sheet and loop-top can be heated to $\sim$10 MK and above in the early impulsive phase \citep[e.g.,][]{2018ApJ...866...64C,2018ApJ...854..122W,2021ApJ...908L..55C,2022ApJ...929...99C} and provide a seed population of electrons with a few keV.
However, we note that this is not common for all flares. Similar results can be obtained using different injection energies.
The TS front in each MHD frame is identified by examining the velocity and Mach number \citep{2018ApJ...869..116S} and 5 keV electrons are injected continuously into the upstream region of the TS.
In each simulation, a total of 9.6 $\times$ 10$^{6}$ pseudo-particles are injected.
To improve the statistics at high energies, we implemented a particle-splitting technique \citep[e.g.,][]{1990ApJ...360..702E,1996JGR...10111095G}, so a pseudo-particle will be split into more particles at higher energies. Particles will be removed from the simulation if it reaches the boundaries of the simulation domain.

\begin{deluxetable*}{lccc}
\tablenum{1}
\tablecaption{Summary of simulation parameters for different runs\label{tab:run}}
\tablewidth{0pt}
\tablehead{
\colhead{Run} & \multicolumn{2}{c}{Turbulent scattering}  & \colhead{Collisional} \\
\colhead{}    & \multicolumn{2}{c}{($D_{\mu \mu 0} \rm, ~  s^{-1}$)}   & \colhead{energy loss} \\
\colhead{}    & \colhead{Looptop}  &  \colhead{Other regions}     & \colhead{}
}
\decimalcolnumbers
\startdata
Run S   &  288    &  288   &  No \\
Run W   &  57.7   &  57.7  &  No \\
Run SW  &  288    &  57.7  &  No \\
Run S-loss  &  288   &  288   &  Yes \\
Run W-loss  &  57.7  &  57.7  &  Yes \\
Run SW-loss &  288   &  57.7  &  Yes \\
\enddata
\tablecomments{Different runs are named with ``S" referring to strong scattering, ``W" weak scattering, ``SW" strong scattering at the loop-top and weak scattering in the rest of the simulation domain. In all simulations, the perpendicular diffusion coefficient is set to be the same, $\kappa_{\perp 0}$ = 2.03 $\times$ 10$^{-4}$ $\kappa_{0}$ = 3.91 $\times$ 10$^{10}$ m$^2$ s$^{-1}$.}
\end{deluxetable*}

\section{Simulation Results} \label{sec:results}
\subsection{Effect of turbulent pitch-angle diffusion on electron acceleration and transport}

\begin{figure}
\centering
\includegraphics[width=0.95\linewidth]{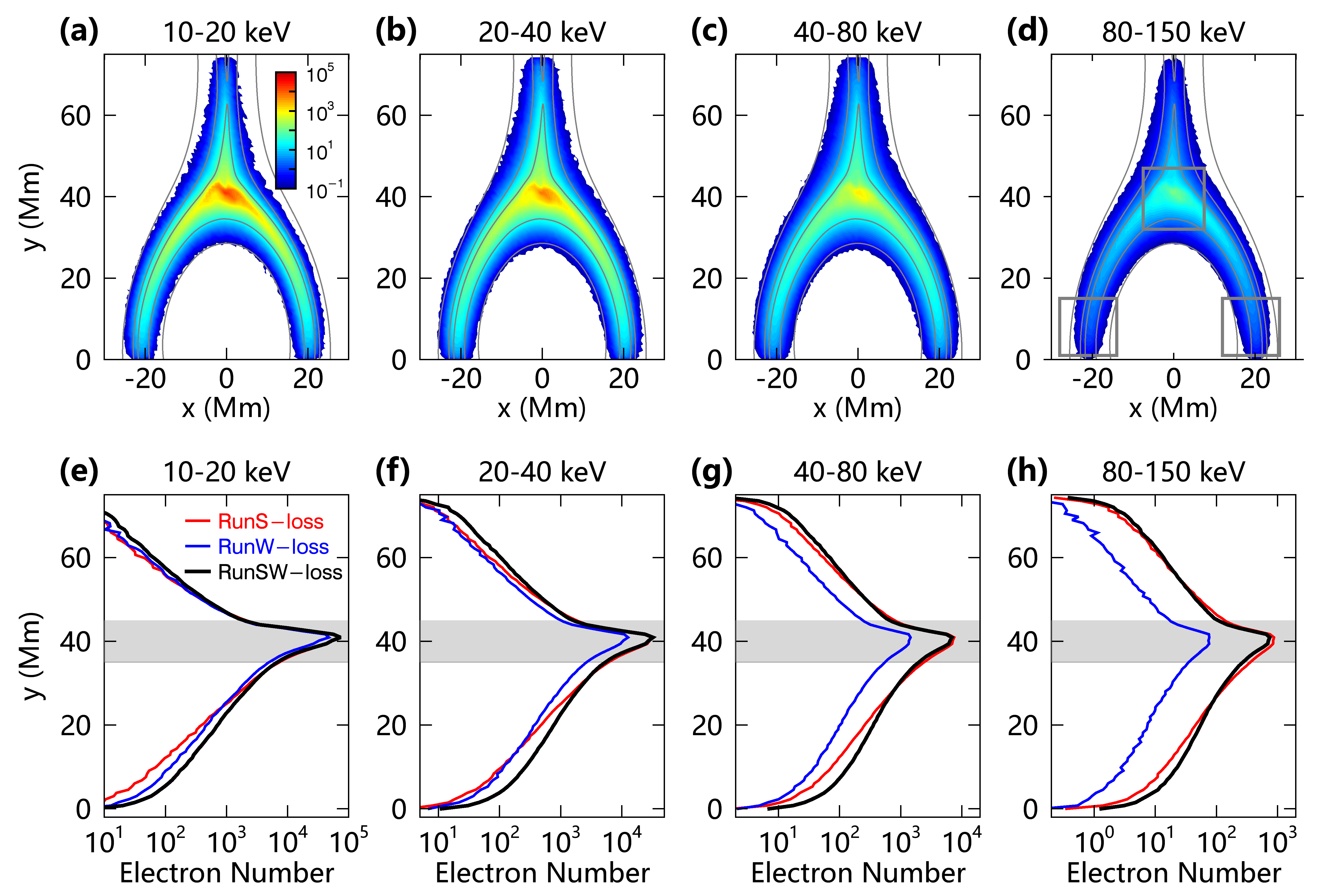}
\caption{
(a)-(d): Spatial distributions of accelerated electrons at different energy ranges from the simulation in Run SW-loss. The color scale in logarithmic  is normalized to the minimum and maximum values as shown in panel (a).
(e)-(h): Distributions of electrons along the $y-$axis (height) after integration over the $x-$axis direction for Run S-loss (red), Run W-loss (blue), and Run SW-loss (black). The gray shaded region between 35-45 Mm indicates the location of the loop-top.
}
\label{fig:dist_ele}
\end{figure}

The overall spatial distributions of accelerated electrons are similar in different simulation runs.
Figures \ref{fig:dist_ele}(a)--(d) show the spatial distributions of accelerated electrons at four different energy ranges from Run SW-loss after a simulation time of $t_0$ = 29.2 s (corresponding to the selected MHD simulation period from 91 $t_0$ to 92 $t_0$).
At all energies, most electrons are concentrated in the loop-top region.
On the one hand, this concentration is due to electron injection and acceleration around the TS in the loop-top region. On the other hand, various effects, including magnetic mirroring \citep{1998ApJ...505..418F}, pitch-angle scattering \citep{2014ApJ...780..176K}, and partially closed field lines acting as a magnetic trap \citep{2019ApJ...887L..37K} can lead to the confinement of energetic electrons at the loop-top.
We also find that the loop-top concentration is non-symmetric and highly time-dependent (not shown here, see e.g., \citet{2020ApJ...905L..16K}), due to the dynamic evolution of background MHD flow and magnetic fields.

We first examine the effect of pitch-angle diffusion due to magnetic turbulence on particle acceleration and transport by comparing the simulation results from Run S-loss and Run W-loss, with $D_{\mu \mu 0}$ being 288 s$^{-1}$ and 57.7 s$^{-1}$, respectively.
To better illustrate the difference in spatial distributions, we integrate the number of electrons over the $x-$axis direction and plot the integrated number distribution along the $y-$axis (height), as shown in Figures \ref{fig:dist_ele}(e)--(h).
The distributions for Run S-loss and Run W-loss are plotted in red and blue, respectively.
In the loop-top ($y \sim$ 35$-$45 Mm, shaded), the number of electrons in Run W-loss is less than that in Run S-loss and the difference gets progressively larger with increasing energy, reaching nearly one order of magnitude at $\sim$80 keV.
However, in the lower portion (legs) of the flare loop, it shows opposite relations at low and high energies.
At low energies (below $\sim$40 keV) there are more electrons in Run W-loss, while at high energies there are much less electrons in Run W-loss.
This indicates that with weaker turbulent scattering in Run W-loss, the low-energy electrons can escape the acceleration site in the loop-top more easily and precipitate into the footpoints.
This is consistent with the modeling result in \citet{2018A&A...610A...6M} that the spatial distribution gets broader and the maximum of the distribution decreases with increasing mean free path.
On the other hand, the acceleration of electrons to higher energies takes a longer time and requires sufficient trapping near the acceleration site.
Weaker turbulent scattering in Run W-loss leads to less efficient acceleration of electrons. As a consequence, the number of high-energy (above $\sim$40 keV) electrons is much smaller in both the loop-top and loop leg regions.
Note that similar effects can be seen in the current sheet region as in the loop legs.
Our simulation results suggest that the impact of turbulent scattering on the spatial distribution of nonthermal electrons, therefore the relative intensity between coronal and footpoint X-ray sources, is highly energy-dependent.

As discussed above, we also consider the case with spatial-dependent turbulent scattering. In Run SW-loss, turbulent scattering in the loop-top where the electrons are mainly accelerated is enhanced. This enables both efficient electron acceleration to high energies in the loop-top and transport of a sufficient number of electrons to the footpoints.
As shown in Figures \ref{fig:dist_ele}(e)--(h), in the loop-top, the number of electrons in Run SW-loss (the curves in black) is close to that in Run S-loss and much larger than that in Run W-loss at energies above 20 keV. In the loop legs, the number of electrons in Run SW-loss is much larger in comparison with both Run S-loss and Run W-loss at all energies.

\begin{figure}
\centering
\includegraphics[width=0.95\linewidth]{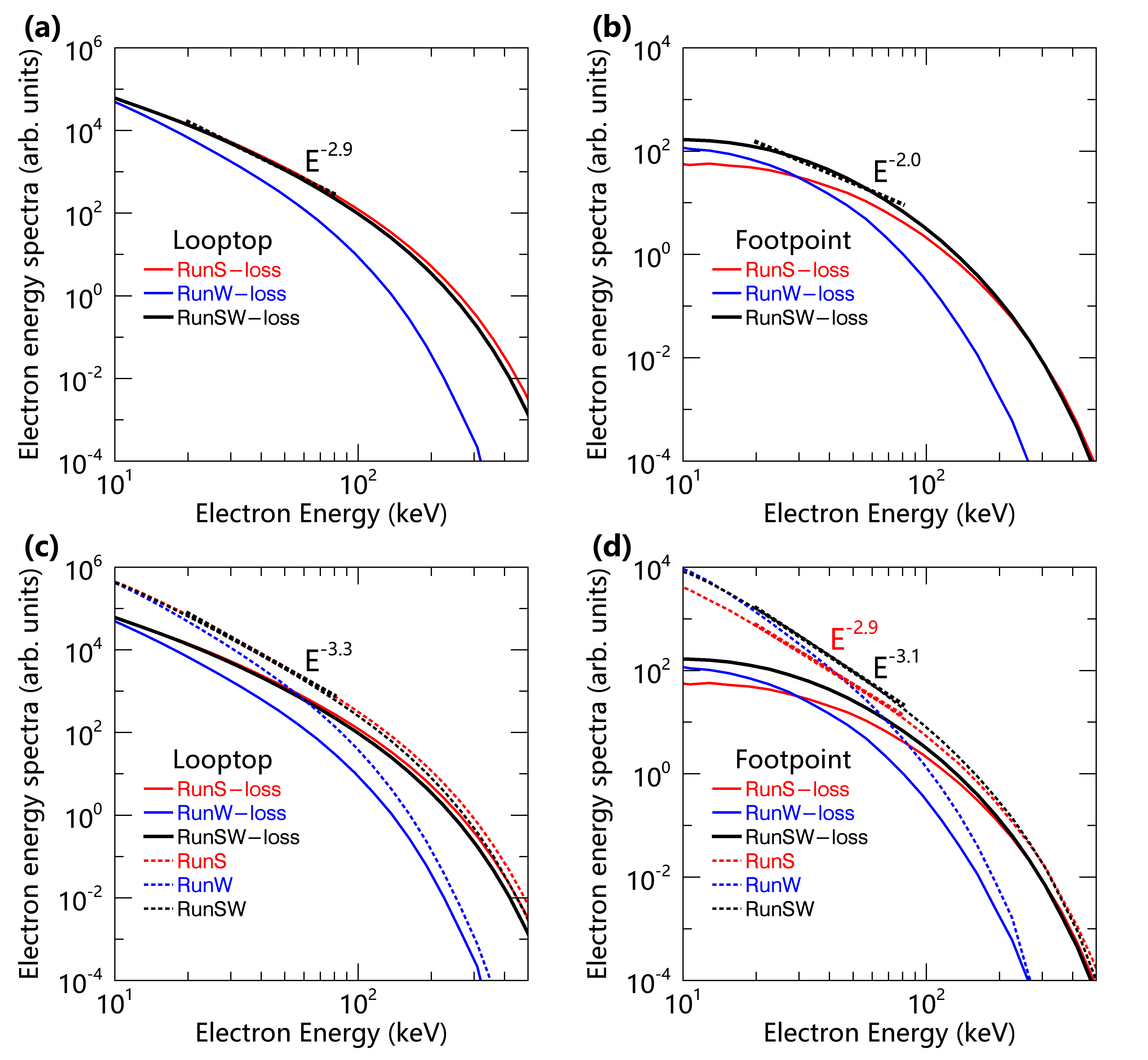}
\caption{
Differential density energy spectra of accelerated electrons integrated over the loop-top (left) and footpoint (right) regions, as indicated by the three gray boxes in Figure 2(d), are plotted in arbitrary units. For the footpoint spectrum, it is the average of the two footpoints. The electron spectra between 20-80 keV are fitted with a single power-law function.
Panels (a) and (b): at the loop-top, the spectral indexes are $\sim$2.9 for Run S-loss and Run SW-loss, while at the footpoints, the spectra flatten at low energies and the spectral index is $\sim$2.0 for Run SW-loss.
The simulation results without collisional loss are displayed in dashed lines in the lower panels for comparison. Panels (c) and (d): at the loop-top, the spectral indexes are $\sim$3.3 for Run S and Run SW, while at the footpoints, the indexes are $\sim$2.9 and $\sim$3.1, respectively.
}
\label{fig:spec_ele}
\end{figure}

The effect of turbulent pitch-angle scattering can also be seen in the energy spectra of accelerated electrons.
Figures \ref{fig:spec_ele} (a) and (b) show the electron differential density spectra, $N(E) \propto pf(p)$, in the loop-top and footpoint regions for the three runs, Run S-loss, Run W-loss, and Run SW-loss.
In Run S-loss and Run SW-loss, with strong scattering at the loop-top, the energy spectrum between 20--80 keV in the loop-top region can be well fitted by a power-law, with a spectral index $\sim$2.9, while the spectrum in the footpoints is relatively harder and rolls over toward lower energies, approximately a power-law with a spectral index $\sim$2.0.
Note that the hardening of electron spectrum between coronal and footpoint sources has also been observed in certain flares \citep{2006A&A...456..751B}, and has been interpreted as preferential energy loss for lower-energy electrons due to Coulomb collisions \citep{2006A&A...456..751B} or return current \citep{2017ApJ...851...78A}.
For comparison, in Figures \ref{fig:spec_ele} (c) and (d), we also display the electron spectra for the three runs without collisional loss in dashed lines.
Without collisonal loss, it shows a weaker hardening at the footpoints, with the power-law spectral index varying from $\sim$3.3 to $\sim$2.9 ($\sim$3.1) for Run S-loss (Run SW-loss).
As assumed in our model, electrons with higher energies have a relatively larger mean free path ($\lambda_{\parallel}$ = $3\kappa_{\parallel} / v \approx 4.86 v^{1/3} L_c^{2/3} \Omega_0^{-1/3} / \sigma^2$), therefore more high-energy electrons would make their way to the footpoints for a given time, resulting in a harder spectrum.
When the effect of collisional loss is added, the hardening of low-energy spectrum is more significant.

In the loop-top region (Figure \ref{fig:spec_ele}(a)), in comparison with Run S-loss, the energy spectrum in Run W-loss gets increasingly steeper at higher energies due to inefficient acceleration, while it is very similar in Run SW-loss.
This implies that scattering enhancement at the loop-top in Run SW-loss causes sufficient acceleration of electrons to high energies as in Run S-loss.
In the footpoints (Figure \ref{fig:spec_ele}(b)), the energy spectrum of Run W-loss intersects that of Run S-loss at $\sim$30 keV, in agreement with the energy-dependent influence of turbulent scattering on the spatial distribution as discussed above.
That is, with weaker scattering, the electrons spend less time at the loop-top acceleration region before escaping to the footpoints, leading to inefficient acceleration of electrons to high energies.
In Run SW-loss, weaker scattering in the flare loop results in more electrons with energy up to $\sim$300 keV precipitating into the footpoints than that in Run S-loss and a softer spectrum.

\subsection{Synthetic HXR emission}
Based on the spatially resolved distributions of energetic electrons and the thermal plasma density from the MHD simulation (Figure \ref{fig:mhd}(c)), we calculate the X-ray emission produced by accelerated electrons.
For each grid from the particle modeling, we calculate the thin-target bremsstrahlung X-ray spectrum by assuming the standard Bethe$-$Heitler cross-section using the Python package \texttt{sunxpsex}, modified to allow calculation using array-based electron distributions from our particle model. In order to produce the synthetic HXR images and spectra, we assume each pseudo-particle count in the simulation represents $10^{6}$ actual electrons per unit volume. This way, the average nonthermal electron density in the looptop region above 10 keV is scaled to $\sim\!7\times 10^8$ cm$^{-3}$, or $\sim$35\% of the background (thermal) plasma density of $\sim\!2\times10^9$ cm$^{-3}$. A rough estimate suggests that this requires about $10\%$ of the shocked electrons to be accelerated. This level of efficiency is supported by recent kinetic simulations \citep{2012ApJ...753...28G,2014ApJ...794..153G,2022ApJ...925...88H}. To simulate the footpoint HXR sources, we place an artificial ``chromosphere'' at the height of $y=7$ Mm. This height is selected to be at a sufficiently large distance (4--5 mean free paths) away from the bottom boundary of the simulation domain near which the number of particles shows a precipitous drop (see, e.g., Figures \ref{fig:dist_ele}(e)--(h)) as they exit the bottom boundary. Meanwhile, this selected height of the chromosphere is low enough to ensure that the magnetic topology at the footpoint region remains nearly unchanged from that of the true bottom boundary. The HXR flux at each grid point at $y=7$ Mm is then calculated based on the thick-target bremsstrahlung scenario. The total electron flux (in electrons s$^{-1}$) reaching each grid point is estimated as $F = v_d n_e^{>E_{\rm min}}A_X$, where $v_d\approx \sqrt{2E_{\rm min}/3m_e}$ is the downward velocity component of the electrons (assuming equipartition; see, e.g., \citealt{2011SSRv..159..225W}), $n_e^{>E_{\rm min}}$ is the total nonthermal electron density above a low-energy cutoff $E_{\rm min}$ taken as 10 keV, and $A_X=dx\times l_z$ is the footpoint area at the grid point with a grid size of $dx$ and column depth of $l_z$. A uniform column depth of $l_z=10$ arcsec is assumed throughout the simulation domain. The resulting X-ray flux calculated using both the thin-target bremsstrahlung (coronal portion) and thick-target X-ray bremsstrahlung (chromosphere portion) is combined to form the final synthetic X-ray images at different photon energies.

\begin{figure}
\centering
\includegraphics[width=0.98\linewidth]{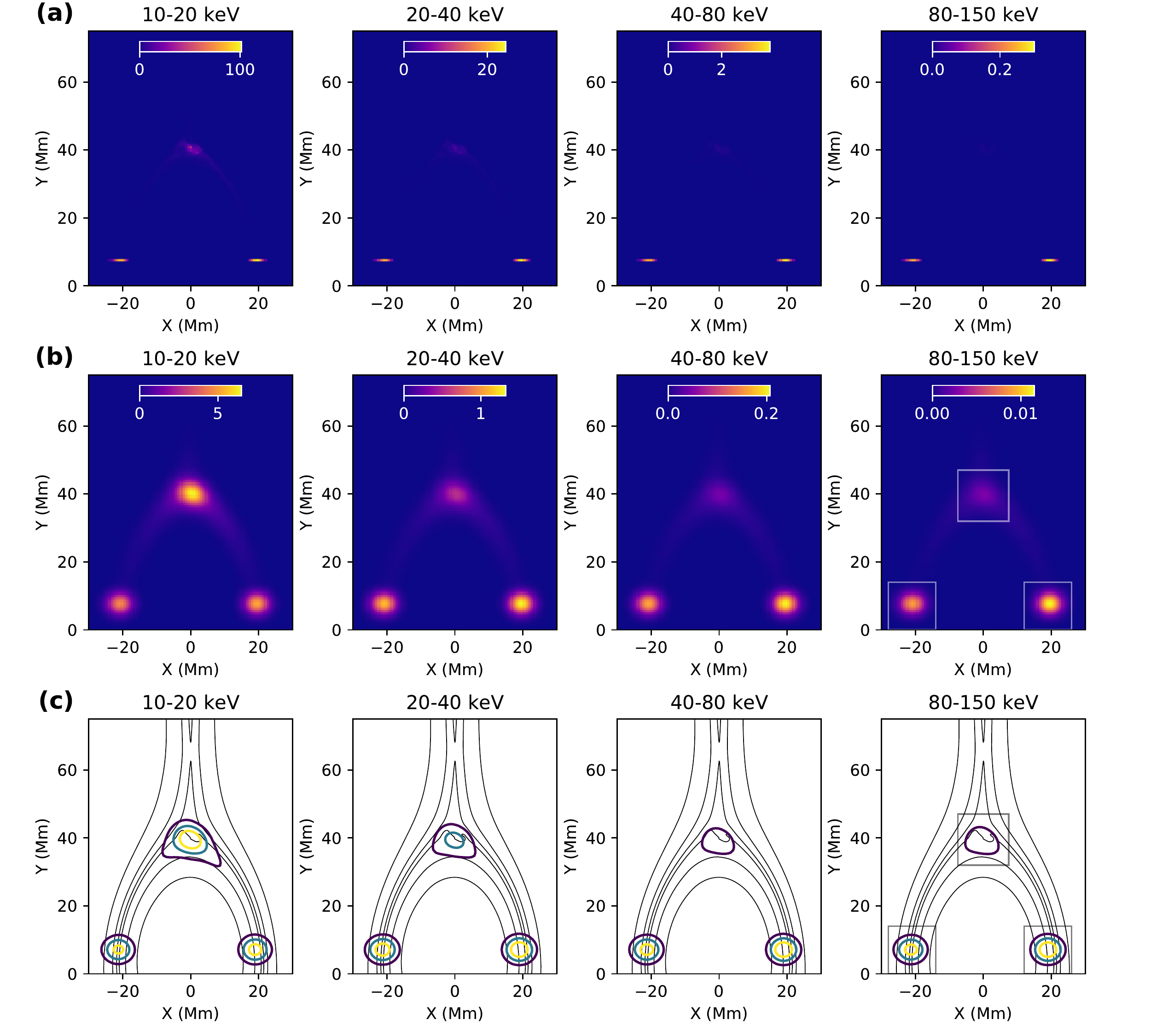}
\caption{
Synthetic HXR images at different photon energy ranges based on the simulation results in Run SW-loss. The color scales are normalized to the minimum and maximum of each individual image. The HXR flux values as observed from Earth (in photons cm$^{-2}$ keV$^{-1}$ s$^{-1}$) are shown in the color bars. (a) Simulated HXR images at the original resolution of the simulation. (b) Simulated HXR images after convolving with a FWHM of 6.8 arcsec (RHESSI detector 3). (c) Same images as in (b) but showing contour levels of 10\%, 30\%, and 60\%. The black curves in the background are selected magnetic field lines.
}
\label{fig:dist_xray_img}
\end{figure}

Figure \ref{fig:dist_xray_img}(a) shows the HXR intensity images at different photon energy ranges based on the simulation results in Run SW-loss at the original resolution of the simulation. Note that the bright footpoint HXR sources are only present at $y=7$ Mm where the thick-target emission occurs. In comparison, the coronal thin-target source is barely visible at high energies. In principle, these synthetic HXR images at different energies can be taken as the input to simulate observables by different HXR instrumentation, provided that the instrument response is known.

To compare our simulated HXR images with typical \textit{RHESSI} observations, we convolve the synthetic images in Figure~\ref{fig:dist_xray_img}(a) with a Gaussian point-spread function with FWHM of 6.8 arcsec, corresponding to the spatial resolution of \textit{RHESSI} detector 3, as shown in Figures \ref{fig:dist_xray_img}(b) and (c).
Note that this simple Gaussian convolution does not take into account RHESSI's full instrument response and details involved in its Fourier-transform-based image deconvolution processes (\citealt{2002SoPh..210...61H, 2002SoPh..210..165S}; see an approach taken by \citealt{2012ApJ...752....4B}).
As expected, the synthetic HXR images display both the loop-top and footpoint sources. Compared to the images at the original resolution, the coronal loop-top source becomes more visible. This is because the brightness of the compact footpoint sources becomes more ``diluted'' due to the coarse instrument angular resolution.
As a result, the apparent brightness of the loop-top source is greater than the footpoint sources at 10--20 keV, but weaker at energies above 20 keV.
A weaker coronal source is consistent with the actual \textit{RHESSI} observations in most flares reported in the literature \citep[e.g.,][]{2006A&A...456..751B,2014ApJ...780..107K}.
In addition, the coronal source gets increasingly weaker than the footpoint sources at higher energies, down to only $\sim$10\% at energies above 40 keV.
As discussed above, since the loop-top source is due to thin-target emission and the footpoint sources are dominated by thick-target emission, it naturally results in a harder spectrum at the footpoints.

\begin{figure}
\centering
\includegraphics[width=0.98\linewidth]{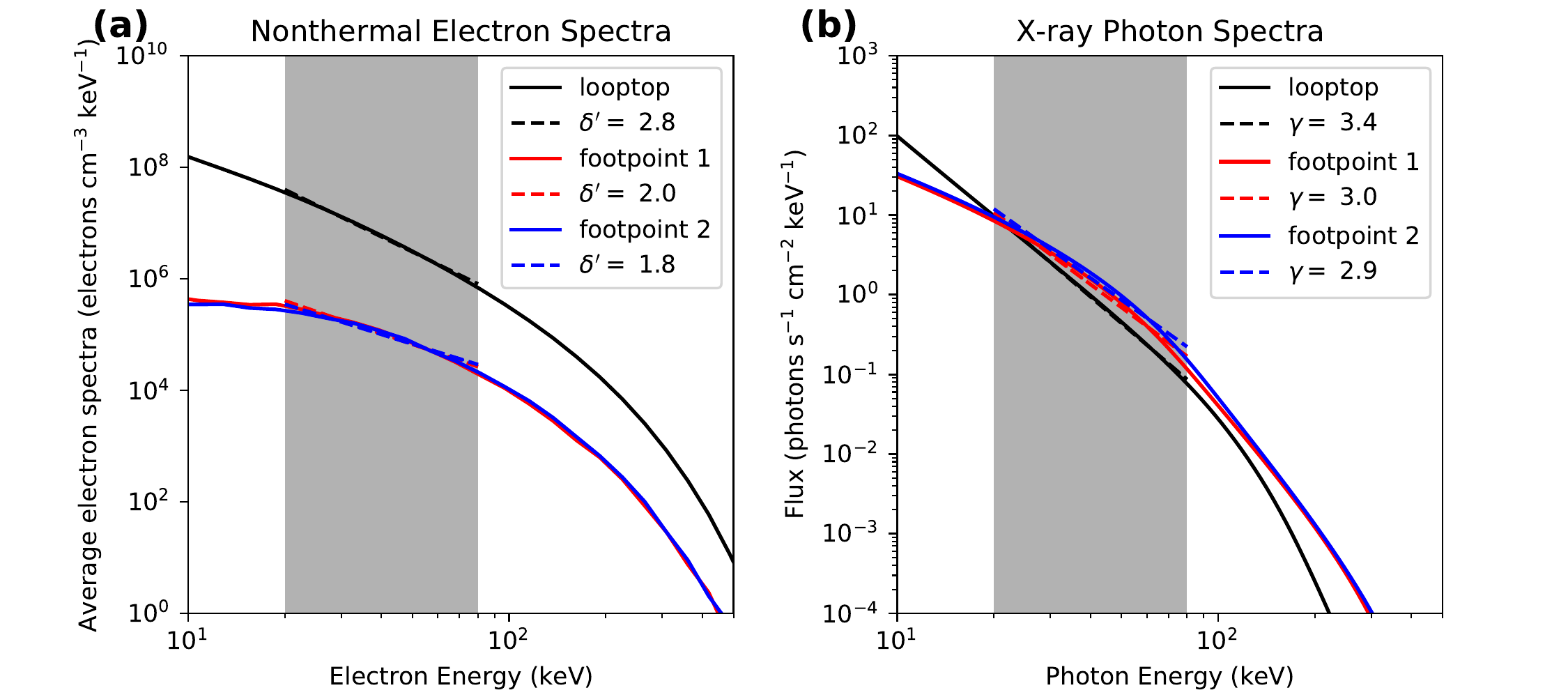}
\caption{
Average electron differential density spectra (a) and X-ray photon flux spectra (b) at the loop-top and two footpoints, as indicated by the three boxes in Figure 4. The spectra between 20-80 keV are fitted with a single power-law function.
}
\label{fig:dist_xray_spec}
\end{figure}

A comparison of the energy spectra between the nonthermal electrons and HXR emission is shown in Figure \ref{fig:dist_xray_spec}.
Figure \ref{fig:dist_xray_spec}(a) shows the average electron differential density spectra (in electrons~cm$^{-3}$~keV$^{-1}$) in the loop-top and two footpoints. For the two footpoints, their average spectra are taken only from the pixels at $y$ = 7 Mm. We fit the spectra between 20--80 keV with a single power-law function, $N(E) \propto E^{-\delta^ \prime}$. The fitted spectral index $\delta^ \prime$ is 2.8 for the loop-top source, and 2.0 and 1.8 for the two footpoints. Therefore, the footpoint electron spectrum is slightly harder than the loop-top electron spectrum, consistent with the results as discussed above (see Figure \ref{fig:spec_ele}).
Figure \ref{fig:dist_xray_spec}(b) shows the HXR photon spectra (in photons~s$^{-1}$~cm$^{-2}$~keV$^{-1}$) by integrating over the loop-top and footpoint regions (indicated by the three boxes in Figure \ref{fig:dist_xray_img}(c), which are large enough to enclose most of the HXR flux). The HXR spectra between 20--80 keV are also fitted with a single power-law function of the form $I(\varepsilon) \propto \varepsilon^{-\gamma}$, with $\gamma$ being 3.4 for the loop-top, and $\sim$3.0 and $\sim$2.9 for the two footpoints, respectively.
In the loop-top, the difference between the HXR photon and nonthermal electron spectral indexes is $\gamma - \delta^ \prime$ = 0.6, close to the relationship $\gamma_{\rm thin}^{\rm spl} = \delta^ \prime +$ 0.5 as predicted by the thin-target model with a single power-law form \citep{1972SoPh...24..414H,2018SSRv..214...82O}. The slight difference may be due to the inhomogeneity within the loop-top region and/or the deviation of the electron spectrum from the single power-law form.
However, for the footpoints, the simulated HXR spectra are much softer than the prediction by the thick-target model with a single power-law form, which expects $\gamma_{\rm thick}^{\rm spl} = \delta^ \prime -$ 1.5  \citep{2018SSRv..214...82O}.
As a result, the difference in the HXR photon spectral index between the loop-top and footpoint sources is much smaller than $\gamma_{\rm thin}^{\rm spl}-\gamma_{\rm thick}^{\rm spl}\approx2$.
The much softer HXR footpoint spectrum in our simulation is due to the rollover of the electron spectrum at higher energies ($\gtrsim$100 keV). Since the X-ray emission at a given photon energy is contributed by the integral of all electrons with energies above it, the relationship between $\gamma_{\rm thick}^{\rm spl}$ and $\delta'$ with the single power-law form is valid only at photon energies one to two orders of magnitude below the break/rollover energy in the electron spectrum \citep{2003ApJ...586..606H,2011SSRv..159..107H}, which is clearly not the case in our simulations where the break energy appears at $\sim$100--200 keV.

As shown in Figure \ref{fig:dist_xray_img}(b), a small difference can be found in the intensity for two footpoint sources, i.e., the right footpoint being slightly brighter at energies above 40 keV. This \red{may be} caused by the asymmetry of the TS structure and magnetic field configuration in the loop-top (see Figure \ref{fig:mhd}), resulting in asymmetric distribution of energetic electrons in the flare loop (Figures \ref{fig:dist_ele}(a)--(d)).
The difference is also visible in the HXR energy spectra.
The discrepancy of spectral indices between two footpoints was found in some flare events and has been explained by effects such as asymmetric magnetic mirroring and column density in the flare loop \citep[e.g.,][]{2006A&A...456..751B,2008SoPh..250...53S,2009ApJ...693..847L}.
Nevertheless, we can not rule out the possibility of numerical fluctuations in the model.

\section{Conclusions and Discussion} \label{sec:conclusion}
In this paper, we present a macroscopic particle model for studying the acceleration and transport of nonthermal electrons in solar flares. We numerically solve the focused particle transport equation by combining it with time-dependent plasma flow and magnetic fields provided by MHD simulations.
Our particle model naturally incorporates the acceleration and transport processes.
In contrast, the electron acceleration process is rarely included in previous flare particle transport models and electron injection in the loop-top is based on simplified assumptions.

In our simulations, electrons are primarily accelerated by the TS and confined in the loop-top by a magnetic bottle structure that features a local minimum in magnetic field strength.
Here we discuss the effects of turbulent pitch-angle scattering on the spatial distribution and energy spectrum of nonthermal electrons in this context.
We find that turbulent scattering can have important impacts on both the electron acceleration in the loop-top and the subsequent transport in the flare loop, and the influences are highly energy dependent.
With weaker turbulent scattering, the low-energy electrons can escape the acceleration site in the loop-top more easily and more electrons can precipitate into the footpoints.
However, because sufficient trapping near the acceleration site is required for producing high-energy electrons, much fewer electrons with energy above $\sim$50 keV are present both in the loop-top and footpoints when the scattering is weak.
Motivated by EUV spectroscopic observations that suggest the presence of a high level of turbulence in the loop-top region \citep[e.g.,][]{2017PhRvL.118o5101K,2021ApJ...923...40S}, we consider a simulation with spatial-dependent turbulent scattering. We show that enhancement of turbulent scattering in the loop-top can enable both efficient electron acceleration to high energies and transport of abundant electrons to the footpoints.
We generate spatially resolved synthetic X-ray emission images and spectra by combining the thin-target bremsstrahlung model for the whole domain with the thick-target model for the footpoints. Both the coronal and footpoint sources can be observed, while the intensity of the coronal source is much weaker and the spectrum is softer compared to the footpoint sources.

The focused transport equation used in our work includes various physics of electron acceleration and transport. Our simulation results are generally consistent with the discussions in previous flare models \citep[e.g.,][]{2013ApJ...777...33C,2018A&A...610A...6M}, particularly at low energies below $\sim$50 keV.
Transport effects, such as Coulomb collisions and return current, can modify the spatial distribution and energy spectrum of nonthermal electrons, therefore are important to understanding the X-ray sources in flare observations \citep[e.g.,][]{2006ApJ...651..553Z,2014ApJ...787...86J,2018SciA....4.2794J}.
These physical processes can be easily included in the Fokker-Planck modeling \citep[e.g.,][]{1998ApJ...505..418F,2014ApJ...780..176K,2020ApJ...902...16A}.
In this work, we have included the effects of collisions on energy loss and pitch-angle scattering, and other transport effects will be discussed in future work.
In addition, the bottom boundary in the MHD simulation will be improved by including the high-density chromosphere \citep[e.g.,][]{2020ApJ...897...64Y}.
Our macroscopic particle model can be used to produce spatially resolved synthetic images of nonthermal emissions in HXRs and, in principle, microwaves, thereby enabling direct comparison with flare observations. We suggest that such practice would have strong implications for further understanding the high-energy particle acceleration and transport processes in solar flares.

According to the standard model of solar flares, a pair of flare ribbons in the chromosphere outline the footpoints of reconnecting magnetic field lines in the current sheet. Since magnetic reconnection in the current sheet is hard to be observed directly, observations of flare ribbons can be used to infer the properties in the reconnecting current sheet, e.g., the reconnection rate, the variation of the guide field.
Recently, it has been found that the $>$25 keV HXR emission correlates well with the dynamical evolution of flare ribbons \citep{2022ApJ...926..218N,2022SoPh..297...80Q}.
This implies a strong correlation between the production of nonthermal electrons and the dynamics of the reconnecting current sheet.
Furthermore, the rapid rise of the nonthermal HXR emission usually takes place a few minutes after the onset of the flare, when the magnetic shear inferred from the evolution of flare ribbon brightenings is decreasing.
In some flares, the onset of hot (10$-$15 MK) soft X-ray emission occurs prior to the detection of any HXR emission \citep{2021MNRAS.501.1273H}.
In future work, combining our macroscopic particle model with the 3D MHD simulation of dynamically evolving reconnecting current sheet in solar flares \citep[e.g.,][]{2022NatAs...6..317S,2022ApJ...932...94D} may be a promising approach to unveil the underlying mechanisms.

\bibliography{export-bibtex}{}
\bibliographystyle{aasjournal}

\end{document}